# Spectral Phase Pulse Shaping Alters Photoionization Time

J. Aygun, D. Yaacoub[†], A. L. Harris*

Department of Physics, Illinois State University, Normal, IL, USA

## Abstract

Photoionization is a key step in many attosecond processes. Accurately determining the photoionization time delay is critical to understanding electron dynamics during and after ionization and can guide future efforts to manipulate electron motion. Prior studies have shown that the photoionization time delay is non-zero and that pulse shaping through alteration of the spectral phase may change the number and timing of ionization events. In order to more quantitatively assess whether and how the spectral phase modifies the photoionization time delay, we use attosecond streaking simulations to extract the streaking time delay for ionizing pulses with identical power spectra, but different spectral phases. We compare streaking time delays for Gaussian, Airy, and fifth order phase extreme ultraviolet (XUV) pulses. We find that the streaking delay depends on the XUV spectral phase and that the sign of the delay is determined by the sign of the phase for large phases. Pulses with non-zero spectral phase show an asymmetry in the streaking spectrogram that is associated with phase-dependent spectral compression or broadening. Comparison of the streaking delays for short- and long-range potentials indicates that Coulomb-laser coupling contributions to the streaking shift are independent of spectral phase, confirming that the observed phase dependence arises from intrinsic photoionization dynamics. Overall, our work suggests that the spectral phase may open the door to new opportunities for controlling ionization timing and provide new avenues for coherent control of ultrafast electron dynamics.

## I. Introduction

Much of attosecond physics focuses on the observation of electron dynamics on their natural time scale. Photoionization is one of the fundamental electron transition processes, occurring in numerous attosecond processes. Over the last several decades, improvements in the production of ultrashort and phase-controlled laser pulses [1–3] have allowed the field to address the fundamental question of how long it takes for an electron to escape its binding potential during photoionization [4]. This time is known as the ionization time delay and it has importance beyond fundamental atomic physics. For example, photoionization underlies many other processes, such as high-order harmonic generation (HHG) and above threshold ionization (ATI), and descriptions of these processes often rely on electron trajectories that directly depend on the so-called birth time of the electron into the continuum. Thus, accurate knowledge of the photoionization time delay is critical to describing and understanding these processes.

Techniques such as attosecond streaking [5] and reconstruction of attosecond harmonic beating by interference of two-photon transitions (RABBITT) [6,7] are often used to measure the photoionization time delays, and most studies now point to non-zero times on the order of tens of attoseconds [4,8–12]. Here, we focus on attosecond streaking and the influence of the spectral

*corresponding author: alharri@ilstu.edu
[†]current address: Department of Physics, University of Toronto, 60 St. George Street, Toronto, ON M5S 1A7, Canada; Canadian Institute of Theoretical Astrophysics, 60 St. George Street, Toronto, Toronto, ON M5S 3H8, Canada

phase on the photoionization time delay. Attosecond streaking is a pump-probe process in which a short extreme ultraviolet (XUV) pulse ionizes the atom in the presence of a weak infrared (IR) field. Following ionization, the electron moves classically in the IR field, where it gains momentum and energy due to the IR field vector potential. The final photoelectron momentum shift (streaking shift) relative to the IR field vector potential depends on the phase of the IR field at the time of ionization. Thus, the ionization time is mapped to the photoelectron momentum. By varying the time delay between the XUV and IR pulses, a streaking spectrum is obtained, and the photoionization time delay can be found.

In addition to improved measurement of the photoionization time delay, the ability to manipulate it could lead to greater control over electron dynamics. To this end, significant progress has been made in the ability to shape ultrashort laser pulses. For example, pulse shaping through the use of 2-color pulses has been shown to yield both atomic structure and electron trajectory information [13]. Alternatively, the introduction of a time-dependent spectral phase (chirp) provides a means to manipulate the instantaneous frequency and temporal localization of the ionizing field. In the case of attosecond streaking, the effects of pulse chirp were found to be negligible when the photoionization cross sections varied smoothly with frequency, but played a significant role for frequencies near the Cooper minimum [14]. Yet another method of pulse shaping can be achieved by controlling the pulse's spectral phase, which, in the case of a third order spectral phase, leads to an Airy temporal pulse profile. In this case, simulations of HHG and ATI using Airy laser pulses have predicted that the third order spectral phase of the Airy pulse alters the number and timing of the ionization events, as well as the timing of recollisions [15,16].

Most attosecond streaking studies to date have focused on extracting delays assuming transform-limited pulses, but given the indications from HHG and ATI studies that changes in spectral phase may alter ionization time delays, it is important to understand what role the spectral phase plays in the ionization time. This information is also essential to determining the extent to which ionization timing can be actively controlled. Here, we use numerical simulations of attosecond streaking to quantify the effect of third (Airy) and fifth order spectral phases on the streaking delay. Unlike pulse chirp, in which the spectral phase of the pulse is time-dependent, the pulses used here have a constant spectral phase. This has the effect of introducing an asymmetry into the temporal envelope of the pulse, but does not alter the power spectrum. Thus, it is possible to directly compare streaking delays for pulses with identical power spectra, but different spectral phases and temporal envelopes.

We present attosecond streaking simulations using the one-dimensional time-dependent Schrödinger equation (TDSE) to determine the effect of the spectral phase on the streaking time delay. We compare results for ionization by XUV pulses with zero spectral phase (Gaussian), third order spectral phases (Airy), and fifth order spectral phases using a short-range Yukawa potential and a long-range pliant core potential. We show that the streaking delay is spectral phase dependent, and that for large spectral phase, the sign of the delay matches the sign of the phase. Additionally, a positive spectral phase causes a spectral compression in the streaking spectrum, while a negative spectral phase leads to a spectral broadening in the streaking spectrum. Comparison of the streaking delays for short- and long-range potentials shows that the coupling between the laser field and the long-range Coulomb potential is not spectral phase dependent.

The remainder of the paper is organized as follows. Section II contains a description of the numerical methods used in the simulations. Section III has results and discussion for attosecond streaking using XUV pulses with different spectral phases. Section IV contains a summary and conclusions. Atomic units are used throughout unless otherwise noted.

## II. Numerical Methods
### A. TDSE

For linearly polarized laser pulses, the primary dynamics occur along the polarization direction. Thus, a one-dimensional model is typically sufficient to capture the relevant physics. We solved the one-dimensional TDSE in the dipole approximation and length gauge for a single active electron atom ionized by an XUV pulse in the presence of a weak IR field

$$i\frac{\partial}{\partial t}\psi(x,t) = \left[-\frac{1}{2}\frac{d^2}{dx^2} + V_a(x) + xE_{XUV}(t) + xE_{IR}(t)\right]\psi(x,t), \tag{1}$$

where $V_a(x)$ is the atomic potential and $E_{XUV,IR}(t)$ is the XUV or IR electric field.

The Crank-Nicolson method [17] was used to solve Eq. (1). The initial state wave function was found by imaginary time propagation, and absorbing boundary conditions [18] were used to prevent reflections from the spatial grid boundary. The density of the wave function was checked at each time step to ensure that no probability was lost. The spatial grid spanned from -6000 a.u. to 6000 a.u. with a step size of 0.07 a.u. A temporal step size of 0.005 a.u. was used.

### B. Ionization Time Delays

The intrinsic photoionization time delay is given by the Eisenbud, Wigner, and Smith (EWS) time [19] $t_{EWS}$. In attosecond streaking, the goal is to determine this intrinsic photoionization time delay from the streaking shift $t_{streak}$, which can be found from fitting the first moment of the streaking spectrum $p_{COM}$ to the time-shifted IR vector potential

$$p_{COM} = p_0 - A_{IR}(\tau + t_{streak}). \tag{2}$$

Here, $p_0$ is the asymptotic momentum of the photoelectron in the absence of a streaking field, $A_{IR}$ is the vector potential of the IR pulse, and $\tau$ is the time delay between the temporal centers of the IR and XUV pulses.

To calculate the streaking spectrum, the initial state wave function was propagated to a final time $t_f$ of twice the pulse duration. This ensured that the photoelectron had travelled sufficiently far from the atomic potential. We then performed a Fourier transform on the spatially separated ionized electron wave packet at $t_f$ to find the momentum distribution [20]. The first moment of the electron momentum was calculated using

$$p_{COM} = \frac{\int_{p_{min}}^{p_{max}} p|\phi(p)|^2 dp}{\int_{p_{min}}^{p_{max}} |\phi(p)|^2 dp}, \tag{3}$$

where $|\phi(p)|^2$ is the momentum density, and the integration was performed in a range encompassing the photoelectron momentum (here, $p_{min,max}$ were 1 and 2 a.u. respectively). $\tau$ was varied over the central cycle of the IR streaking field and the resulting streaking spectrum showed the expected sinusoidal behavior with a shift of $t_{streak}$ from the vector potential of the IR pulse.

For short-range potentials and symmetric ionizing pulses, the streaking shift is simply equal to the EWS time

$$t_{streak} = t_{EWS}. \tag{4}$$

For long-range potentials, the EWS time and the streaking shift are not equal, but rather the streaking shift is a sum of two terms: a Coulomb analog of the EWS time $t_{EWS}^C$ and a term due to the Coulomb distortion of the wave front $\Delta t_{Coul}$

$$t_{streak} = t_{EWS}^C + \Delta t_{Coul}. \tag{5}$$

The first term is the intrinsic delay for photoionization from a long-range potential and the second term accounts for the slowdown of the electron in the Coulomb field, also known as the Coulomb-laser coupling term. As in the case of a short-range potential, $t_{streak}$ can be found by fitting the first moment to the time-shifted vector potential.

### C. Pulses

A 6-cycle Gaussian IR streaking pulse was used with an intensity of $I = 10^{12}$ W/cm² ($E_{0IR} = 0.0053$ a.u.), wavelength of 1000 nm ($\omega_{IR} = 0.0456$ a.u.) and full-width half maximum of $dt = 275$ a.u. (6.66 fs). Its electric field is given by

$$E_{IR}(t) = E_{0IR} e^{-2\ln 2 (t-t_c)^2/dt^2} \sin(\omega_{IR}(t - t_c)), \tag{6}$$

where $E_{0IR}$ is the maximum electric field strength, $t_c$ is the temporal center of the pulse, and $\omega_{IR}$ is the central frequency.

Three different XUV ionizing pulse types were used in order to examine the effect of the spectral phase. In all cases, the XUV pulse had 20 cycles, an intensity of $I = 10^{13}$ W/cm² ($E_{0XUV} = 0.017$ a.u.), photon energy of 55 eV ($\omega_{XUV} = 2$ a.u.), and full-width half maximum of $dt = 12.6$ a.u. (305 as). The power spectrum of all XUV pulses was kept identical, but the phases were different. A Gaussian pulse has zero spectral phase and in the frequency domain is given by

$$E_G(\omega) = \frac{E_0 dt \sqrt{\pi}}{2^{3/2}\sqrt{\ln 2}} e^{-\frac{dt^2(\omega_0 - \omega)^2}{(8\ln 2)}}. \tag{7}$$

In the time domain, the functional form of the Gaussian pulse is identical to that of Eq. (6).

An Airy pulse has a third order spectral phase $\phi_3$ and in the frequency domain is given by [21]

$$E_A(\omega) = E_G(\omega) e^{-\frac{i}{6}\phi_3(\omega-\omega_0)^3}. \tag{8}$$

In the time domain, the electric field of the Airy pulse is given by

$$E_A(t) = E_0 \sqrt{\frac{\pi}{2\ln 2}} \frac{dt}{\tau_0} Ai\left(\frac{\tau-(t-t_c)}{\Delta\tau}\right) e^{\frac{\ln 2\left(\frac{2\tau}{3}-(t-t_c)\right)}{2\tau_{1/2}}} \sin(\omega_0(t-t_c)), \quad (9)$$

where $\Delta\tau$ and $\tau_{1/2}$ are the stretch and truncation half-life of the pulse, respectively. Higher order spectral phases can be introduced into the electric field of the XUV pulse, and we additionally performed calculations for pulses with a fifth order spectral phase $\phi_5$

$$E_A(\omega) = E_G(\omega) e^{-i\phi_5(\omega-\omega_0)^5}. \quad (10)$$

The corresponding temporal electric field was found numerically from the inverse Fourier transform of Eq. (10).

For a given temporal center $t_C$ of the XUV pulse, the maximum of the pulse shifts away from $t_c$ as the magnitude of the spectral phase increases. This can alter the streaking spectrum because the streaking shift is defined relative to the peak position of the XUV pulse. To ensure that our calculations of the streaking shift are not affected by the spectral phase dependence of $t_C$, we added an additional shift to the temporal pulses with non-zero spectral phases to ensure that the pulse maxima occurred at the same time, regardless of spectral phase (see Fig. 1).

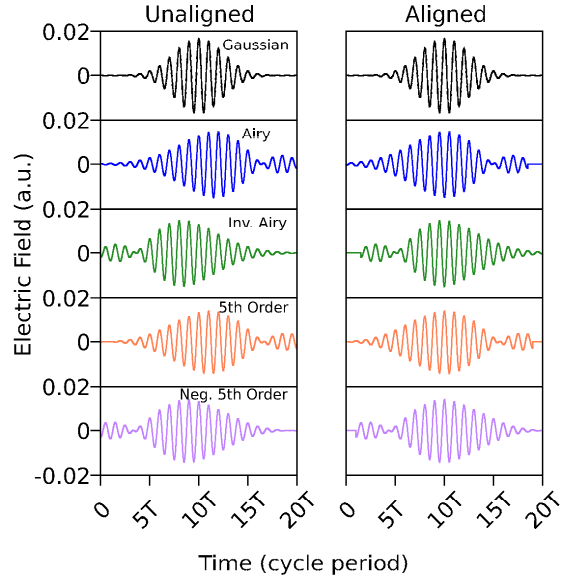

Figure 1 XUV electric fields for different third order and fifth order spectral phases. The left column shows pulses with identical temporal centers $t_c$, and it can be seen that the peak values of the electric fields are shifted from $t_c$ for non-zero spectral phases. The right column shows the same pulses shifted in time to align the maxima of the envelopes. The phase was was $|\phi_3| = 1000$ a.u. for the Airy and inverted Airy pulses and $|\phi_5| = 5000$ a.u. for the 5th order and negative 5th order pulses shown here.

### D. Atomic Potentials

We performed calculations for both short- and long-range potentials that model a hydrogen atom. For the short-range potential, we used a Yukawa potential given by [22]

$$V_a(x) = -\frac{e^{-\frac{|x|}{10}}}{\sqrt{x^2+1.2741^2}}. \tag{11}$$

The parameters used in the Yukawa potential ensure that the lowest eigenstate has a binding energy matching that of hydrogen (13.6 eV).

For the long-range potential, we used a Pliant core potential [23] that has been shown to reasonably approximate the magnitude and plateau cutoff values of 3-dimensional HHG and ATI calculations [23,24]

$$V_a(x) = -\frac{1}{\left(|x|^{\frac{3}{2}}+1.45\right)^{2/3}}. \tag{12}$$

Unlike the more commonly used soft-core potential, the pliant core potential has a sharper cusp feature at the origin that more accurately represents the atomic potential near the nucleus, while still resembling the Coulomb potential at asymptotic distances. The parameters used in this pliant core model potential yield an ionization energy of 13.6 eV.

## III. Results
### A. Short-range potentials

For short-range potentials, the streaking shift is the intrinsic time delay for ionization and can be found directly from the first moment of the streaking spectrum. Physically, it represents the delay of the ionization time relative to the peak of the XUV pulse. To determine the effect of the spectral phase on the streaking shift, we solved the one-dimensional TDSE for ionization of a 1s electron from a Yukawa potential by an XUV pulse in a weak IR streaking field. The XUV pulse's spectral phase was either zero (Gaussian), cubic (Airy), or quintic (fifth order). All XUV pulses were designed such that they had identical power spectra. For the Airy pulses, the magnitude of the phase was varied in the range $100 \leq |\phi_3| \leq 1000$ and for the 5$^{th}$ order phase pulses the range of phases was $100 \leq |\phi_5| \leq 5000$. Figure 2(a-e) shows the streaking spectra (photoelectron momentum as a function of time delay between the peaks of the XUV and IR pulses) for Gaussian, Airy, and 5$^{th}$ order phase pulses with the most extreme phases of $|\phi_3| = 1000$ a.u. and $|\phi_5| = 5000$ a.u. The first moment is shown as the white trace in panels (a-e), and the location of its minimum relative to the minimum of the IR vector potential yields the streaking shift. Figure 2f shows a close-up of the first moments around their minima, and it is clear that the streaking shift is dependent upon the spectral phase, indicating that the spectral phase alters the intrinsic ionization delay time.

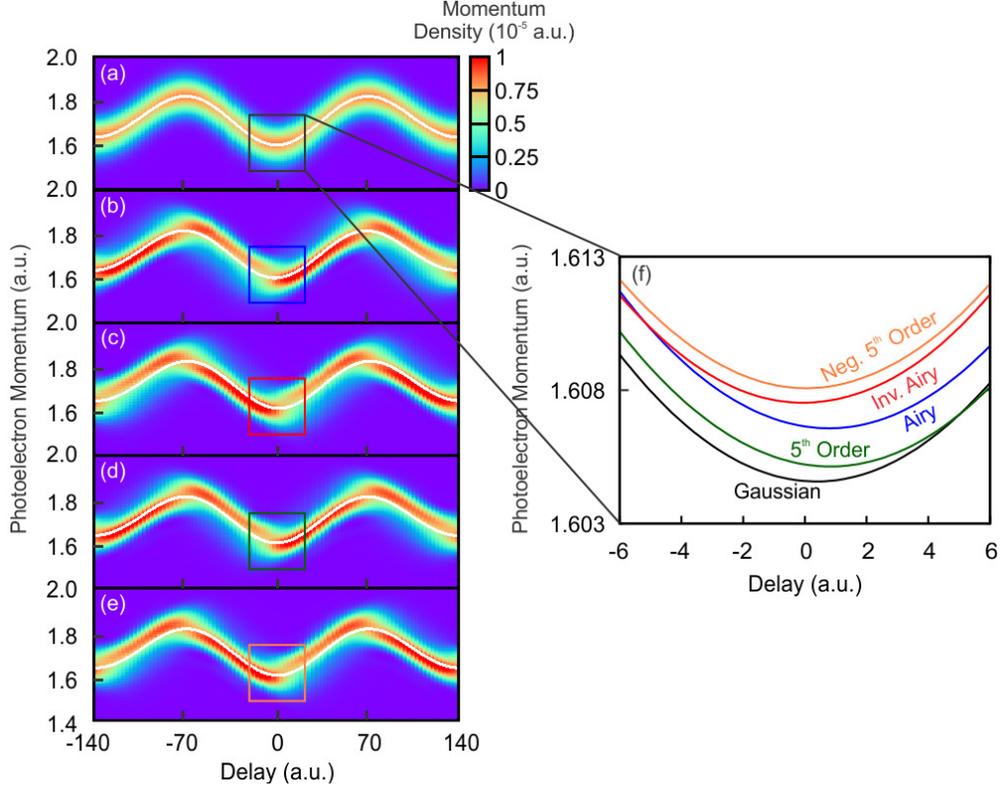

Figure 2 Photoemission streaking spectra for XUV photoionization of a 1s electron from a hydrogen Yukawa potential (Eq. (11)) using a Gaussian pulse (a), an Airy pulse (b; $\phi_3 = 1000$ a.u.), an inverted Airy pulse (c; $\phi_3 = -1000$ a.u.), a 5$^{th}$ order pulse (d; $\phi_5 = 5000$ a.u.), and a negative 5$^{th}$ order pulse (e; $\phi_5 = -5000$ a.u.). The first moment is shown as the white trace in panels (a-e). Panel (f) shows the first moments of the streaking spectra from (a-e) near their minima. Colors in panel (f) correspond to Black – Gaussian pulse; Blue – Airy pulse; Red – inverted Airy pulse; Green – 5$^{th}$ order pulse; Orange – negative 5$^{th}$ order pulse.

Figure 3 shows the dependence of the streaking shift on the spectral phase as found from Eq. (2). For both 3$^{rd}$ order and 5$^{th}$ order spectral phases, this dependence is approximately sigmoidal with a larger magnitude of the delay for larger absolute phases. A Gaussian pulse has zero spectral phase and results in a positive streaking shift of 0.5 a.u. As the magnitude of the phase increases, the streaking shift approaches fixed values for both positive and negative phases, indicating that it becomes essentially independent of the spectral phase. With a few exceptions for small negative phase values, the sign of the delay is the same as the sign of the phase.

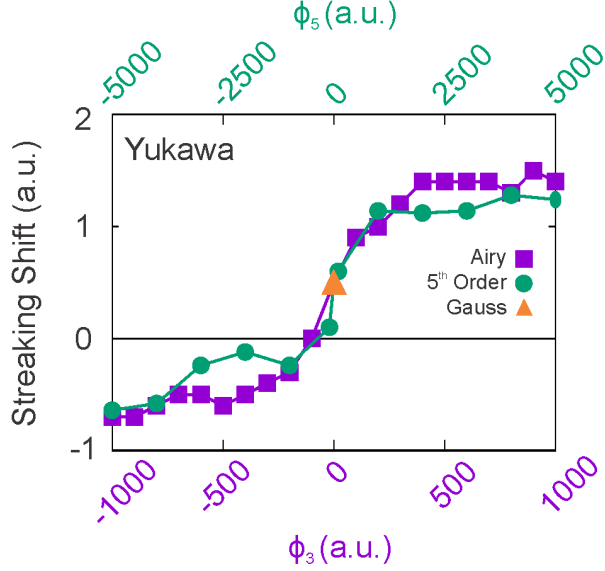

Figure 3 Streaking shift ($t_{streak}$) from a hydrogen Yukawa potential as a function of spectral phase for Airy (purple squares) and 5$^{th}$ order (green circles) phase XUV pulses. The time delay for a Gaussian pulse is shown as an orange triangle. Connecting lines are drawn to guide the eye.

To further understand why the presence of a non-zero spectral phase in the ionizing pulse changes the streaking shift, we identify several possible physical sources. First, as can be seen from Fig. 1, as the spectral phase changes, the peak of the electric field relative to the peak of the envelope shifts, indicating a change in the carrier envelope phase (CEP) of the pulse. To test whether this change in CEP can account for the change in streaking shift, we introduced a CEP term $\delta$ in the Gaussian pulse

$$E_{IR}(t) = E_{0IR} e^{-2\ln 2 (t-t_c)^2/dt^2} \sin(\omega_{IR}(t-t_c) + \delta). \qquad (13)$$

The CEP for the Gaussian pulse could then be chosen such that the time difference between the maximum of the electric field and the maximum of the envelope for the Gaussian pulse was identical to that of an Airy pulse with $\phi_3 = 1000$ or -1000 a.u. For an Airy pulse with $\phi_3 = 1000$ a.u., the temporal maximum of the envelope occurred at 418.3 a.u., and the maximum of the electric field occurred at 418.5 a.u. This same time difference could be achieved using a Gaussian pulse with $\delta = 0.38\pi$. In these cases, the streaking shift was found to be 1.4 a.u. for the Airy pulse and 0.4 a.u. for the CEP-shifted Gaussian pulse. Likewise, for an Airy pulse with $\phi_3 = -1000$ a.u., the temporal maximum of the envelope occurred at 407.7 a.u. and the maximum of the electric field was at 407.5 a.u. The streaking shifts in these cases were found to be -0.7 a.u. for the inverted Airy pulse and 0.3 a.u. for the CEP-shifted Gaussian pulse. These results indicate that Gaussian and Airy pulses with identical CEPs do not result in the same streaking shifts, and thus the dependence of streaking shift on spectral phase is not simply due to an alteration in CEP.

In addition to altering the CEP, the spectral phase of the XUV pulse alters the pulse width and magnitude. To test whether the increased width of pulses with non-zero spectral phase had an effect on the streaking shift, we performed a calculation using a Gaussian pulse with $dt = 18$ a.u., which more closely resembled the width of the Airy pulse with $\phi_3 = 1000$ a.u. This wider Gaussian pulse resulted in a streaking shift of 0.4 a.u., which is nearly identical to the streaking shift for the more narrow Gaussian pulse and different from the Airy streaking shift, indicating that pulse width was not responsible for the change in streaking shift observed with changing spectral phase. To test the effect of pulse magnitude on the streaking shift, we performed a calculation using an Airy pulse with $\phi_3 = 1000$ a.u. and an intensity of $I = 1.3 \times 10^{13}$ W/cm$^2$. This resulted in the maximum electric field magnitude matching that of the Gaussian pulse with $I = 10^{13}$ W/cm$^2$. This increased intensity had no effect on the streaking shift for the Airy pulse, indicating that the pulse intensity was not responsible for the change in the streaking shift observed as spectral phase changes.

Lastly, a non-zero spectral phase introduces an asymmetry in the pulse envelope. Comparison of the streaking shifts for Airy and 5$^{th}$ order pulses shows similar effects when the pulses have similar asymmetric temporal envelopes. An Airy pulse with $\phi_3 = 1000$ a.u. has a similar pulse envelope shape to a 5$^{th}$ order pulse with $\phi_5 = 5000$ a.u. and the spectral shifts are similar (1.4 a.u. for Airy and 1.24 for 5$^{th}$ order), indicating that the temporal asymmetry is correlated with the streaking shift. To further test the influence of the temporal asymmetry on the streaking shift, we performed a calculation using an XUV pulse with a skew Gauss envelope. The parameters were chosen such that the envelope's shape closely resembled that of the Airy pulse with $\phi_3 = 1000$ a.u.

$$E_{skew}(t) = 1.6 E_0 e^{-2\ln 2 \left(\frac{(t-t_c)}{18}\right)^2} \left(1 + \text{erf}\left(\frac{0.5(t-t_c-18)}{18}\right)\right) \sin(\omega(t-t_c)). \tag{14}$$

In this case, the spectral density of the skew Gauss pulse is not perfectly Gaussian, but nearly so, and is compressed relative to that of the Airy pulse. For the skew Gauss pulse, the streaking shift was 0.3 a.u., which is quite different than that of the Airy pulse (1.4 a.u.). Combined, our calculations for pulses with asymmetric envelopes indicate that pulses with similar temporal asymmetries and identical spectral densities result in similar streaking shifts. However, pulses with similar temporal asymmetries but different spectral densities result in different streaking shifts.

In addition to the change in streaking shift as spectral phase changes, the streaking spectra in Fig. 2(a-e) show an asymmetry in the momentum distribution of the photoelectron. For pulses with positive spectral phases, increased intensity is observed when the vector potential is rising (i.e., $0 \leq \tau \leq 70$) compared to when it is falling (i.e., $-70 \leq \tau \leq 0$). The opposite is true for pulses with negative spectral phases. Closer inspection of the spectrogram in these regions reveals that the photoelectron spectrum is compressed when the vector potential is rising and broadened when the vector potential is falling. This indicates that for pulses with positive spectral phases, a spectral compression occurs when the electric field is positive, while for negative spectral phases, a spectral broadening occurs when the electric field is negative. Such asymmetry in the streaking spectrum is reminiscent of a similar spectral broadening or compression observed in streaking measurements for XUV pulses with chirped pulses [25] and in the spectral width of high order harmonics produced by a chirped field [3]. In the streaking measurements of [25], a positive electric field also led to a compression of the photoelectron spectrum, and thus an enhancement in intensity observed in the streaking spectrum. Likewise, a negative electric field led to a broadening of the photoelectron spectrum and a corresponding reduction in the streaking spectrum intensity. In the case of a chirped XUV pulse, the asymmetric time dependence of the pulse's frequency distribution led to an asymmetric energy distribution of the released photoelectrons and the observed asymmetry in the streaking spectrum. In our case, the XUV pulse's frequency distribution is symmetric and identical for all pulses used. However, the presence of a non-zero spectral phase in the ionizing XUV pulse causes a temporal asymmetry in the pulse envelope, leading to the observed asymmetry in the photoelectron streaking spectrum.

## B. Long-range potentials

Our results for the short-range Yukawa potential showed that a non-zero spectral phase changed the streaking shift. To determine if this was also the case for long-range potentials, we repeated our calculations using the pliant core potential of Eq. (12), and Fig. 4 shows the streaking shift as a function of spectral phase. Again, a clear sigmoidal dependence of the streaking shift on $\phi_3$ and $\phi_5$ was observed with negative phases typically resulting in negative streaking shifts.

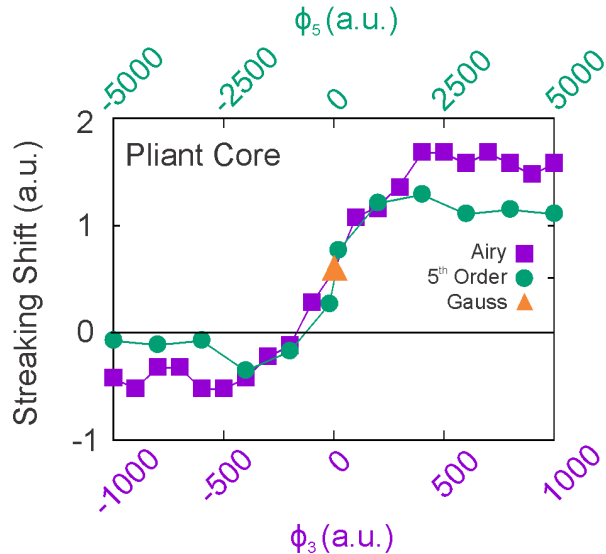

Figure 4 Same as Fig. 3, but for a pliant core potential.

The use of a long-range potential introduces an additional time delay term to the streaking shift due to the Coulomb laser coupling term $\Delta t_{Coul}$ (see Eq. (5)). By taking the difference between the streaking shifts for the pliant core and Yukawa potentials, the dependence of the Coulomb laser coupling term on the spectral phase can be found. This subtraction yields a nearly constant value of $\Delta t_{Coul}$ for both the Airy pulses and the 5th order phase pulses. In particular, for Airy pulses, the average value of $\Delta t_{Coul}$ over $\phi_3$ was 0.2 a.u. with a linear fit slope of $10^{-5}$ a.u. For 5th order phase pulses, the average value of $\Delta t_{Coul}$ over $\phi_5$ was 0.13 a.u. with a linear fit slope of -5 x $10^{-5}$ a.u. The small value of the linear fit slope as either $\phi_3$ or $\phi_5$ changes indicates that the Coulomb laser coupling term is independent of the ionizing pulse's spectral phase.

## IV. Conclusion

We have presented results for attosecond streaking using XUV ionizing pulses with non-zero spectral phases. By solving the time-dependent Schrödinger equation for both short-range and long-range model potentials, we showed that the streaking shift exhibits a clear and systematic dependence on the spectral phase of the ionizing pulse, even when the spectral density is held fixed. For short-range potentials, where the streaking shift directly corresponds to the intrinsic photoionization time delay, our results demonstrated that the ionization time is not solely determined by the energy content of the pulse but is sensitive to its spectral phase. For both third order (Airy) and fifth order spectral phases, the streaking shift displayed a sigmoidal dependence on the magnitude and sign of the phase, with saturation at large absolute phase values. With the exception of small negative spectral phase values, the sign of the streaking shift matched the sign of the spectral phase. Through additional simulations, we eliminated changes in carrier-envelope phase, pulse duration, and peak intensity as the mechanisms responsible for the phase dependence of the streaking shift. Rather, comparison of simulations using Airy and skewed Gaussian pulses showed that the interplay between the spectral phase and the temporal asymmetry of the ionizing pulse is likely responsible for the changes observed in the streaking shift. This indicated that the streaking shift depends on the full spectral phase structure of the ionizing pulse rather than on any single temporal characteristic. Additionally, we observed an asymmetry in the streaking spectrograms for pulses with non-zero spectral phase and traced this to phase-dependent spectral compression or broadening correlated with the sign of the streaking field.

Comparison of streaking shifts for long- and short-range potentials allowed us to isolate the Coulomb-laser coupling contribution to the streaking shift, and our results showed that it is independent of the spectral phase of the ionizing pulse. This indicated that the observed phase dependence of the streaking shift resulted from intrinsic ionization dynamics rather than continuum propagation effects.

Combined, our results demonstrate that attosecond streaking time shifts are sensitive to the spectral phase of the ionizing XUV pulse, even in the absence of changes to the spectral density. These findings have important implications for the interpretation of attosecond time-delay measurements, as they confirm prior results that indicated changes to ionization times in HHG and ATI processes when Airy ionizing pulses were used. Additionally, our work suggests that the spectral phase may provide an additional control parameter for manipulating ionization timing, opening new possibilities for coherent control of ultrafast electron dynamics.

**Acknowledgements**
We gratefully acknowledge the support of the National Science Foundation under Grant No. PHY-2207209, the use of Illinois State University High Performance Computing resources, and additional financial support through a Firebird grant provided by the Office of Student Research at Illinois State University.